\documentclass[aps,prl,twocolumn,groupedaddress,floatfix]{revtex4}
\usepackage{graphicx}
\usepackage{epsfig}
\usepackage{amsmath}

\begin{document}
\title{Magnetocaloric effect in Gd/W thin film heterostructures\\
J. Appl. Phys. 107, 09A903 (2010)}

\author{Casey~W.~Miller}
\affiliation{Department of Physics, Center for Integrated Functional Materials,
University of South Florida, 4202 East Fowler Avenue, Tampa, FL 33620, USA}
\author{D. V. Williams}
\affiliation{Department of Physics, Center for Integrated Functional Materials,
University of South Florida, 4202 East Fowler Avenue, Tampa, FL 33620, USA}
\author{N. S. Bingham}
\affiliation{Department of Physics, Center for Integrated Functional Materials,
University of South Florida, 4202 East Fowler Avenue, Tampa, FL 33620, USA}
\author{H. Srikanth}
\affiliation{Department of Physics, Center for Integrated Functional Materials,
University of South Florida, 4202 East Fowler Avenue, Tampa, FL 33620, USA}

\begin{abstract}
In an effort to understand the impact of nanostructuring on the magnetocaloric effect, we have grown and studied gadolinium in MgO/W(50~$\textrm{\AA}$)/[Gd(400~$\textrm{\AA}$)/W(50~$\textrm{\AA}$)]$_8$  heterostructures.  The entropy change associated with the second order magnetic phase transition was determined from the isothermal magnetization for numerous temperatures and the appropriate Maxwell relation. The entropy change peaks at a temperature of 284~K with a value of approximately 3.4~J/kg-K for a 0-30~kOe field change; the full width at half max of the entropy change peak is about 70~K, which is significantly wider than that of bulk Gd under similar conditions. The relative cooling power of this nanoscale system is about 240~J/kg, somewhat lower than that of bulk Gd (410~J/kg).  An iterative Kovel-Fisher method was used to determine the critical exponents governing the phase transition to be $\beta=0.51$, and $\gamma=1.75$.  Along with a suppressed Curie temperature relative to the bulk, the fact that the convergent value of $\gamma$ is that predicted by the 2-D Ising model may suggest that finite size effects play an important role in this system.  Together, these observations suggest that nanostructuring may be a promising route to tailoring the magnetocaloric response of materials.

\end{abstract}
 \maketitle

\section{Introduction}
Materials exhibiting the giant magnetocaloric effect (MCE) have
demonstrated potential for advancing magnetic refrigeration, an
energy-efficient and environmentally friendly alternative to
conventional refrigeration
\cite{1999JAP....85.5365G,MCEreview.ARMS}. Historically, magnetic refrigerants have been used only for very
specific applications at very low temperatures, and in most cases,
the entropy changes were small \cite{1999JMMM..200...44P}. In
fact, the largest values of the MCE were encountered in the
vicinity of a second-order magnetic transition in Gd
\cite{brown:3673}.  The discovery of the giant MCE in
Gd$_5$(Si$_{1-x}$Ge$_x$)$_4$ has opened the door for highly
efficient magnetic refrigeration near room temperature \cite{PhysRevLett.78.4494,pecharsky:3299}. However,
with the currently available magnetic materials, this high
efficiency is only realized in high magnetic fields. Nanostructuring is a promising route to perturb properties, and may lead to novel and advantageous magnetocaloric properties using existing materials
\cite{mcmichael:29}.  As Gd is traditionally the standard by which all magnetocaloric materials are measured, we have begun investigating the impact of nanostructuring using Gd thin films, specifically in Gd/W thin film heterostructures.

\section{Experimental}

Magnetron sputtering was used with ultra high purity (UHP, 99.999\%) argon gas to deposit W(50~$\textrm{\AA}$)/[Gd(400~$\textrm{\AA}$)/W(50~$\textrm{\AA}$)]$_8$ multilayers onto MgO (100) substrates at ambient temperature in an all-stainless system whose base pressure is 20~nTorr.  Prior to deposition, the MgO substrate surface was modestly milled with 50~W rf biasing in 10~mTorr UHP Ar for ten minutes, and the sputtering targets, 99.95\% W and 99.95\% (rare earth equivalent) Gd, were presputtered under deposition conditions for ten minutes.  In addition to cleaning the target surface, presputtering the Gd serves to getter background gases from the chamber, causing the base pressure to fall by an order of magnitude.  The deposition parameters were 3~mTorr UHP Ar flowing at 20 standard cubic centimeters per minute (SCCM) and 100~W of dc power for W, 100~W rf for Gd. Rate calibrations of samples grown under these conditions were performed on thicker films using x-ray reflectivity; the deposition rates were 0.71~$\textrm{\AA/s}$ for W and 1.20~$\textrm{\AA/s}$ for Gd.  The sample was \textit{in situ} annealed at 600~$^{\circ}$C for one hour after deposition was complete, which is known to improve interface quality \cite{PhysRevB.50.6457}.  X-ray diffraction for $2\theta$ in the range 20-70$^\circ$ reveals and strong Gd (10$\bar{1}$0) and (0001) peaks  (higher order peaks of these are also present).  A simple Scherrer grain size analysis indicates that the Gd layers are structurally coherent throughout their thickness.

\begin{figure}[t]
\begin{center}
\includegraphics[width=3.3in]{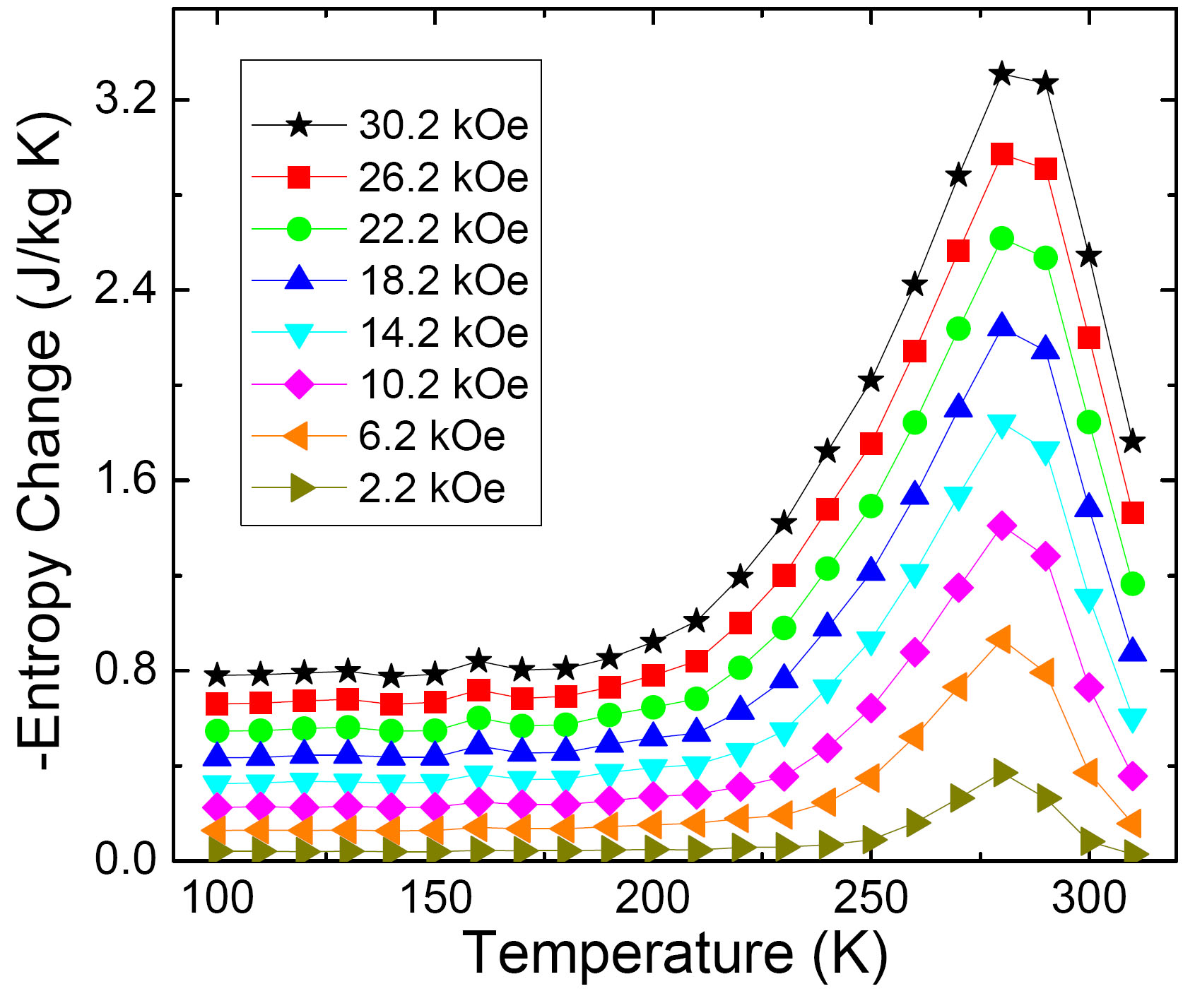}
     \caption{\small{(Color online) Magnetic entropy change of W(50$\textrm{\AA}$)/[Gd(400$\textrm{\AA}$)/W(50$\textrm{\AA}$)]$_8$.
     }} \label{entropy}
\end{center}
\end{figure}

Magnetic measurements were performed using a Quantum Design Physical Property Measurement System. The magnetization isotherms were measured in the range of 0--3~T for temperatures of 260--320~K in steps of 10~K.
The change in the entropy ($S$) of a magnetic material in an applied magnetic field ($H$) is related to the change in magnetization ($M$) with respect to the temperature ($T$) through the thermodynamic Maxwell relation:
\begin{eqnarray*}
\left(\frac{\partial S(T,H)}{\partial H}\right)_T
=\left(\frac{\partial M(T,H)}{\partial T}\right)_H.
	\label{max}
\end{eqnarray*}
For data taken at discrete field and temperature intervals, the change in magnetic entropy, $\Delta S_M$, due to an applied field from 0 to $H_0$ can be approximated as:
\small
\begin{eqnarray*}
\Delta S_M(T,H_o) = \mu_o\sum_i\frac{M_{i+1}(T_{i+1},H) - M_i(T_i,H)}{T_{i+1}-T_i}\Delta H,
	\label{deltaS}
\end{eqnarray*}
\normalsize
where $\mu_o$ is the permeability of free space. Figure~\ref{entropy} shows the magnetic entropy change measured for the W(50~$\textrm{\AA}$)/[Gd(400~$\textrm{\AA}$)/W(50~$\textrm{\AA}$)]$_8$ multilayer (sample area was 1~cm$^2$).  The $\Delta S_m$ peak for low fields is around 284~K, consistent with $T_c$ as determined by the Kovel-Fisher method below \cite{peaknote}.  The magnitude of the peak is about 3.4~J/kg-K, about one third of the value for bulk Gd \cite{phan-mce-review}. The shift of the peak with maximum field is approximately linear in the field range studied, with a slope of about 0.16~K/kOe.

Perhaps the most striking feature of the temperature dependence of $\Delta S_m$ is that its temperature full width at half max $T_{FWHM}$ is nearly double that of bulk Gd under similar field conditions. In fact, the $T_{FWHM}$ observed for $\Delta H =$ 30~kOe exceeds that of bulk Gd for $\Delta H =$ 50~kOe.  While this is compensated for by a reduction of the $\Delta S_m$ peak value, the relative cooling power calculated as $\Delta S_m \times T_{FWHM}$ \cite{phan-mce-review} is 240~J/kg, which is on the order of that for bulk Gd (410~J/kg).  This may indicate that nanostructuring is a potential route for developing magnetic refrigerants with large useful temperature ranges.

\begin{figure}[t]
\begin{center}
\includegraphics[width=3.3in]{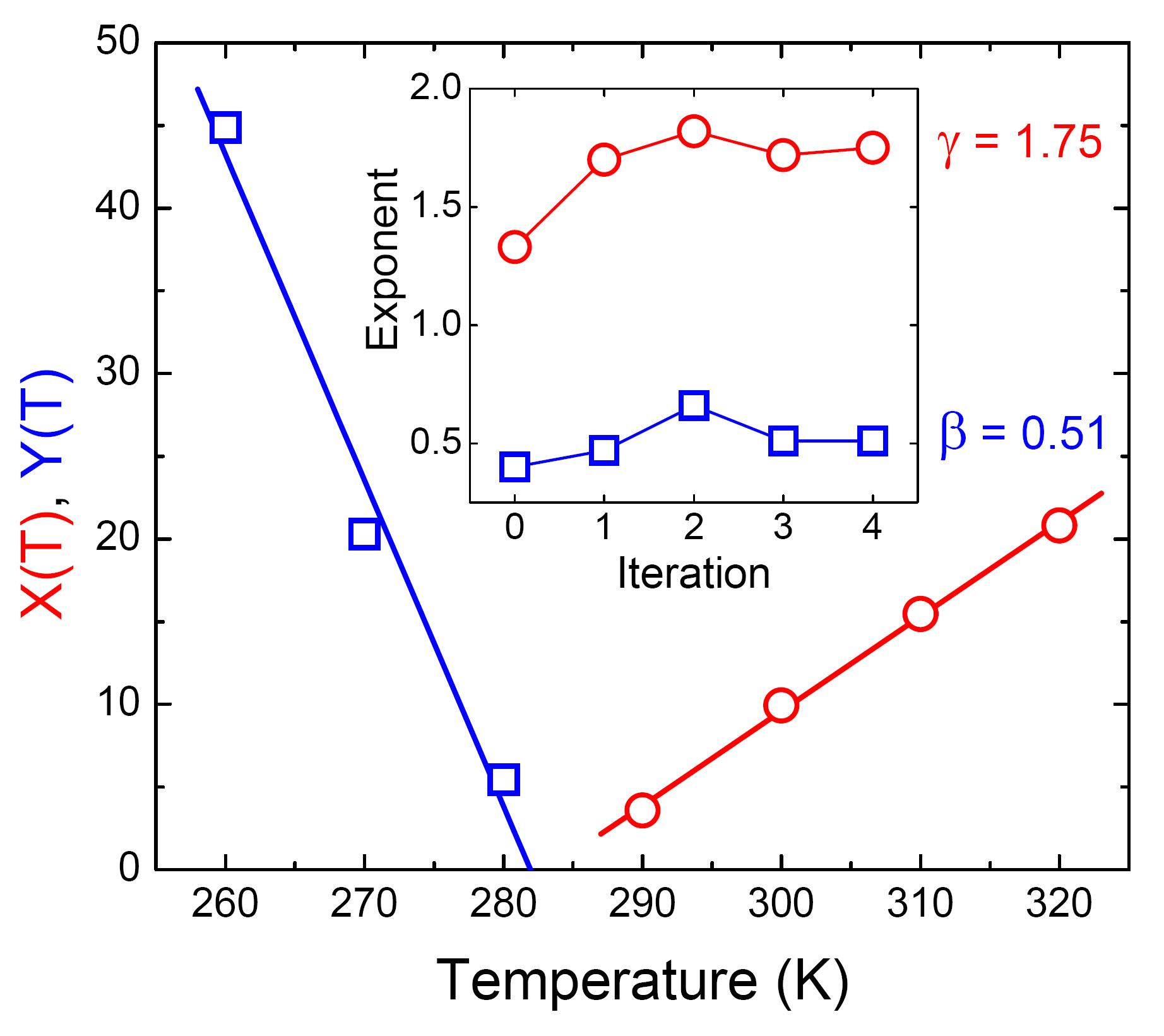}
     \caption{\small{(Color online) Final Y(T) (blue, squares) and X(T) (red, circles) used to determine the convergent critical exponents via the Kovel-Fisher method. The slopes of Y and X are respectively $\beta^{-1}$ and $\gamma^{-1}$. The inset shows the evolution of the critical exponents through four iterations.
     }} \label{exponents}
\end{center}
\end{figure}

\begin{figure}[t]
\begin{center}
\includegraphics[width=3.3in]{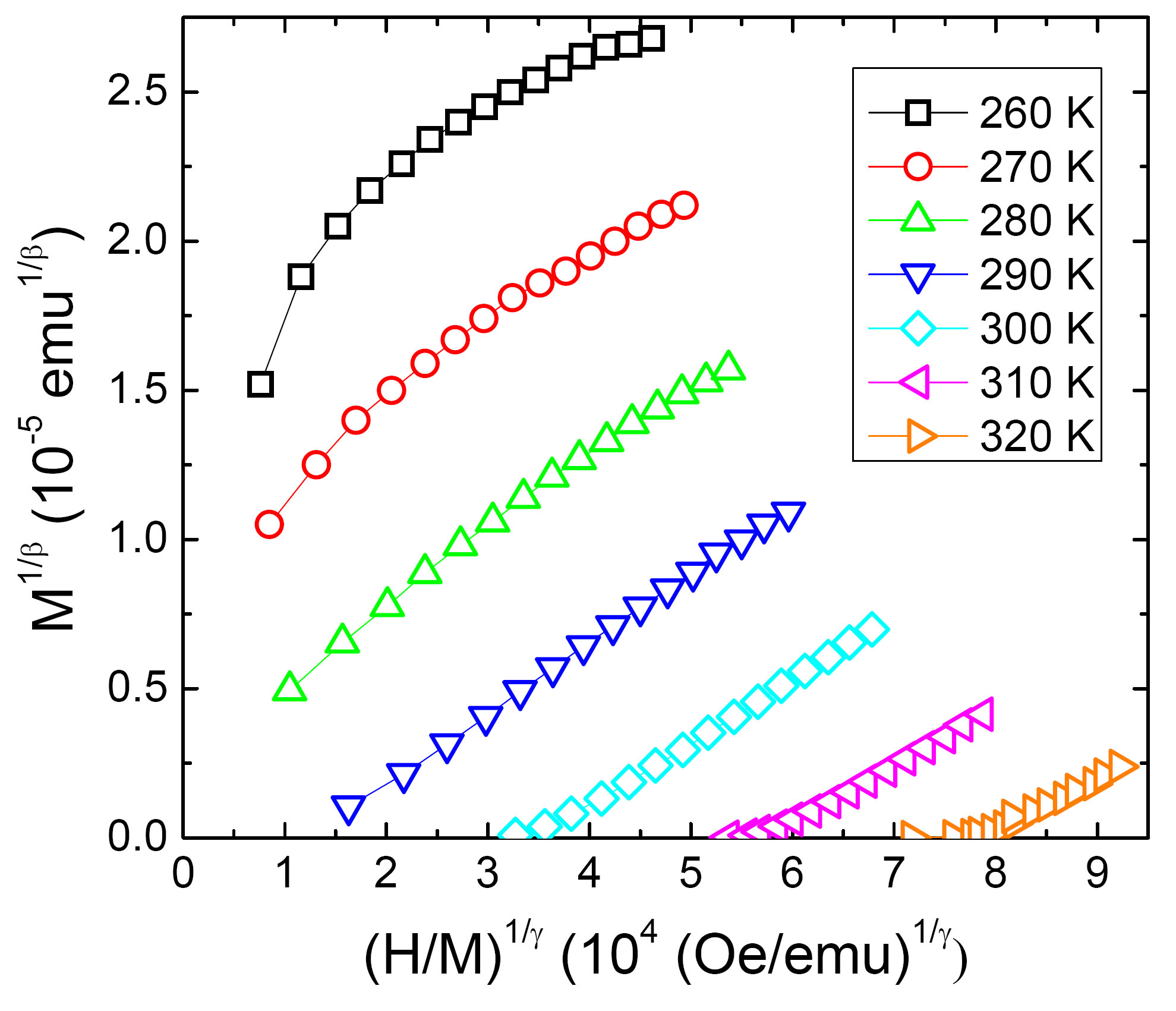}
     \caption{\small{(Color online) Modified Arrott Plot using the critical exponents converged upon after four iterations of the Kouvel-Fisher method: $\beta = 0.51, \gamma = 1.75$.
     }} \label{makf}
\end{center}
\end{figure}

An iterative Kouvel-Fisher method was used to determine the critical exponents and Curie temperature of the system \cite{PhysRev.136.A1626}.  This approach has been quite successful for analyzing amorphous magnetic materials \cite{franco:033903,0953-8984-20-28-285207}. The data were initially analyzed on a modified Arrott-Noakes plot, $M^{1/\beta_o}~vs~(H/M)^{1/\gamma_o}$, where $\beta_o=2/5$ and $\gamma_o=4/3$ \cite{PhysRevLett.19.786}. The high field portions ($H>1~T$) of each isothermal data set were extrapolated to determine the $M^{2.5}$ and $(H/M)^{0.75}$ intercepts with second order polynomials \cite{Kaul19855}, which allows us to determine the spontaneous magnetization ($M_s$) and the inverse of the initial susceptibility ($\chi^{-1}_o$), respectively.  The slopes of the functions $Y(T) = M_s/(dM_s/dT)$ and $X(T) = \chi^{-1}_o/(d\chi^{-1}_o/dT)$ near $T_c$ are taken as $1/\beta$ and $1/\gamma$, respectively.  The initial exponents, $\beta_o=2/5$ and $\gamma_o=4/3$, led to nonlinear $Y(T)$ and $X(T)$; second order polynomials fit these functions well, and were used to estimate $\beta_1$ and $\gamma_1$ near $T_c$. These exponents were then used to create a new modified Arrott-Noakes plot ($M^{1/\beta_1}~vs~(H/M)^{1/\gamma_1}$), thus beginning another iteration of this process.  The data were analyzed in this fashion for four iterations at which point the exponents were unchanging ($\beta_3 = \beta_4$, and $\gamma_3 = \gamma_4$).  The convergent exponents were determined to be $\beta=0.51$, and $\gamma=1.75$. Figure~\ref{exponents} shows that the final $Y(T)$ and $X(T)$ obtained by following this procedure are linear on either side of the second order phase transition; the inset shows the values of the two exponents after each iteration.  Interestingly, the Curie temperature was relatively insensitive to this iterative process, ranging from 282~K to 284~K for each iteration. This Curie temperature is reduced from that of bulk Gd, which is consistent with finite size effects in Gd thin films \cite{jiang:5615}. Figure~\ref{makf} shows the final modified Arrott-Noakes plot using the exponents $\beta=0.51$, and $\gamma=1.75$.

\section{Discussion}The origin of the enhanced $T_{FWHM}$ of the entropy change is currently under investigation.  The critical exponent $\gamma$ determined above to be 1.75 is consistent with the two-dimensional Ising model, potentially suggesting the origin may be related to a change in dimensionality \cite{PhysRevB.57.2701}.  The suppressed Curie temperature of the Gd does indicate conclusively that finite size effects are playing a significant role in the system, further lending some credence to this possibility.  Another potential origin of the enhanced $T_{FWHM}$, as well as the reduced entropy change peak value, is a distribution of Curie temperatures within the Gd. Such a situation could arise, for instance, if the Curie temperature were suppressed near the Gd/W interfaces.  Indeed, the interfacial interaction between Fe and Gd in Fe/Gd superlattices causes the Gd moments near the interfaces to exhibit ferromagnetic order above the bulk Curie temperature and with a magnetic moment exceeding that of bulk Gd \cite{PhysRevB.70.134420}. It is therefore conceivable that the Curie temperature would be suppressed at the Gd-W interface.  An investigation of this sort will likely require depth profiling by polarized neutron reflectometry \cite{kirby:07C929,kirby:245304}.

\section{Conclusion}

We have investigated Gd/W multilayers in the context of the magnetocaloric effect.  It is clear that nanostructuring Gd significantly impacts the behavior of this material.  Relative to bulk Gd, 400~$\textrm{\AA}$ thick Gd films have a reduced entropy change peak value and enhanced entropy change full width at half maximum.  A reduced Curie temperature and susceptibility exponent $\gamma$ of 1.75 suggest that finite size effects are impacting the transition, and may be related to the departure of the magnetic entropy change behavior in thin films from that of bulk material.

\begin{acknowledgments}
Supported by the AFOSR Young Investigator Program (CWM); DOE-BES grant number DE-FG02-07ER46438DVW (HS); the NSF Florida Georgia Louis Stokes Alliance for Minority Participation Bridge to the Doctorate Award, NSF-HRD \#0217675 (DVW); the Center for Integrated Functional Materials is supported by the USAMRMC.
\end{acknowledgments}

\begin{thebibliography}{20}
\expandafter\ifx\csname natexlab\endcsname\relax\def\natexlab#1{#1}\fi
\expandafter\ifx\csname bibnamefont\endcsname\relax
  \def\bibnamefont#1{#1}\fi
\expandafter\ifx\csname bibfnamefont\endcsname\relax
  \def\bibfnamefont#1{#1}\fi
\expandafter\ifx\csname citenamefont\endcsname\relax
  \def\citenamefont#1{#1}\fi
\expandafter\ifx\csname url\endcsname\relax
  \def\url#1{\texttt{#1}}\fi
\expandafter\ifx\csname urlprefix\endcsname\relax\def\urlprefix{URL }\fi
\providecommand{\bibinfo}[2]{#2}
\providecommand{\eprint}[2][]{\url{#2}}

\bibitem[{\citenamefont{{Gschneidner} and
  {Pecharsky}}(1999)}]{1999JAP....85.5365G}
\bibinfo{author}{\bibfnamefont{K.~A.} \bibnamefont{{Gschneidner}}}
  \bibnamefont{and} \bibinfo{author}{\bibfnamefont{V.~K.}
  \bibnamefont{{Pecharsky}}}, \bibinfo{journal}{J. Appl. Phys.}
  \textbf{\bibinfo{volume}{85}}, \bibinfo{pages}{5365} (\bibinfo{year}{1999}).

\bibitem[{\citenamefont{{Gschneidner} and {Pecharsky}}(2000)}]{MCEreview.ARMS}
\bibinfo{author}{\bibfnamefont{K.~A.} \bibnamefont{{Gschneidner}}}
  \bibnamefont{and} \bibinfo{author}{\bibfnamefont{V.~K.}
  \bibnamefont{{Pecharsky}}}, \bibinfo{journal}{Annu. Rev. Mater. Sci.}
  \textbf{\bibinfo{volume}{30}}, \bibinfo{pages}{387} (\bibinfo{year}{2000}).

\bibitem[{\citenamefont{{Pecharsky} and
  {Gschneidner}}(1999)}]{1999JMMM..200...44P}
\bibinfo{author}{\bibfnamefont{V.~K.} \bibnamefont{{Pecharsky}}}
  \bibnamefont{and} \bibinfo{author}{\bibfnamefont{K.~A.}
  \bibnamefont{{Gschneidner}}, \bibfnamefont{Jr.}}, \bibinfo{journal}{J. Magn.
  Magn. Mater.} \textbf{\bibinfo{volume}{200}}, \bibinfo{pages}{44}
  (\bibinfo{year}{1999}).

\bibitem[{\citenamefont{Brown}(1976)}]{brown:3673}
\bibinfo{author}{\bibfnamefont{G.~V.} \bibnamefont{Brown}},
  \bibinfo{journal}{J. Appl. Phys.} \textbf{\bibinfo{volume}{47}},
  \bibinfo{pages}{3673} (\bibinfo{year}{1976}).

\bibitem[{\citenamefont{Pecharsky and Gschneidner}(1997)}]{PhysRevLett.78.4494}
\bibinfo{author}{\bibfnamefont{V.~K.} \bibnamefont{Pecharsky}}
  \bibnamefont{and} \bibinfo{author}{\bibfnamefont{K.~A.}
  \bibnamefont{Gschneidner}, \bibfnamefont{Jr.}}, \bibinfo{journal}{Phys. Rev.
  Lett.} \textbf{\bibinfo{volume}{78}}, \bibinfo{pages}{4494}
  (\bibinfo{year}{1997}).

\bibitem[{\citenamefont{Pecharsky and {Gschneidner,
  Jr.}}(1997)}]{pecharsky:3299}
\bibinfo{author}{\bibfnamefont{V.~K.} \bibnamefont{Pecharsky}}
  \bibnamefont{and} \bibinfo{author}{\bibfnamefont{K.~A.}
  \bibnamefont{{Gschneidner, Jr.}}}, \bibinfo{journal}{Appl. Phys. Lett.}
  \textbf{\bibinfo{volume}{70}}, \bibinfo{pages}{3299} (\bibinfo{year}{1997}).

\bibitem[{\citenamefont{Mcmichael et~al.}(1992)\citenamefont{Mcmichael, Shull,
  Swartzendruber, Bennett, and Watson}}]{mcmichael:29}
\bibinfo{author}{\bibfnamefont{R.~D.} \bibnamefont{Mcmichael}},
  \bibinfo{author}{\bibfnamefont{R.~D.} \bibnamefont{Shull}},
  \bibinfo{author}{\bibfnamefont{L.~J.} \bibnamefont{Swartzendruber}},
  \bibinfo{author}{\bibfnamefont{L.~H.} \bibnamefont{Bennett}},
  \bibnamefont{and} \bibinfo{author}{\bibfnamefont{R.}~\bibnamefont{Watson}},
  \bibinfo{journal}{J. Magn. Magn. Mater.} \textbf{\bibinfo{volume}{111}},
  \bibinfo{pages}{29} (\bibinfo{year}{1992}).

\bibitem[{\citenamefont{Pang et~al.}(1994)\citenamefont{Pang, Berger, and
  Hopster}}]{PhysRevB.50.6457}
\bibinfo{author}{\bibfnamefont{A.~W.} \bibnamefont{Pang}},
  \bibinfo{author}{\bibfnamefont{A.}~\bibnamefont{Berger}}, \bibnamefont{and}
  \bibinfo{author}{\bibfnamefont{H.}~\bibnamefont{Hopster}},
  \bibinfo{journal}{Phys. Rev. B} \textbf{\bibinfo{volume}{50}},
  \bibinfo{pages}{6457} (\bibinfo{year}{1994}).

\bibitem[{pea()}]{peaknote}
\bibinfo{note}{Peak positions were determined by fits to an asymmetric double
  sigmoidal function, which yielded $\chi^2$ values less than unity for all
  data sets.}

\bibitem[{\citenamefont{{Phan} and {Yu}}(2007)}]{phan-mce-review}
\bibinfo{author}{\bibfnamefont{M.-H.} \bibnamefont{{Phan}}} \bibnamefont{and}
  \bibinfo{author}{\bibfnamefont{S.-C.} \bibnamefont{{Yu}}},
  \bibinfo{journal}{J. Magn. Magn. Mater.} \textbf{\bibinfo{volume}{308}},
  \bibinfo{pages}{325} (\bibinfo{year}{2007}).

\bibitem[{\citenamefont{Kouvel and Fisher}(1964)}]{PhysRev.136.A1626}
\bibinfo{author}{\bibfnamefont{J.~S.} \bibnamefont{Kouvel}} \bibnamefont{and}
  \bibinfo{author}{\bibfnamefont{M.~E.} \bibnamefont{Fisher}},
  \bibinfo{journal}{Phys. Rev.} \textbf{\bibinfo{volume}{136}},
  \bibinfo{pages}{A1626} (\bibinfo{year}{1964}).

\bibitem[{\citenamefont{Franco et~al.}(2008{\natexlab{a}})\citenamefont{Franco,
  Conde, and Kiss}}]{franco:033903}
\bibinfo{author}{\bibfnamefont{V.}~\bibnamefont{Franco}},
  \bibinfo{author}{\bibfnamefont{A.}~\bibnamefont{Conde}}, \bibnamefont{and}
  \bibinfo{author}{\bibfnamefont{L.~F.} \bibnamefont{Kiss}},
  \bibinfo{journal}{J. Appl. Phys.} \textbf{\bibinfo{volume}{104}},
  \bibinfo{eid}{033903} (\bibinfo{year}{2008}{\natexlab{a}}).

\bibitem[{\citenamefont{Franco et~al.}(2008{\natexlab{b}})\citenamefont{Franco,
  Conde, Romero-Enrique, and Blazquez}}]{0953-8984-20-28-285207}
\bibinfo{author}{\bibfnamefont{V.}~\bibnamefont{Franco}},
  \bibinfo{author}{\bibfnamefont{A.}~\bibnamefont{Conde}},
  \bibinfo{author}{\bibfnamefont{J.~M.} \bibnamefont{Romero-Enrique}},
  \bibnamefont{and} \bibinfo{author}{\bibfnamefont{J.~S.}
  \bibnamefont{Blazquez}}, \bibinfo{journal}{J. Phys. Condens. Matter.}
  \textbf{\bibinfo{volume}{20}}, \bibinfo{pages}{285207}
  (\bibinfo{year}{2008}{\natexlab{b}}).

\bibitem[{\citenamefont{Arrott and Noakes}(1967)}]{PhysRevLett.19.786}
\bibinfo{author}{\bibfnamefont{A.}~\bibnamefont{Arrott}} \bibnamefont{and}
  \bibinfo{author}{\bibfnamefont{J.~E.} \bibnamefont{Noakes}},
  \bibinfo{journal}{Phys. Rev. Lett.} \textbf{\bibinfo{volume}{19}},
  \bibinfo{pages}{786} (\bibinfo{year}{1967}).

\bibitem[{\citenamefont{Kaul}(1985)}]{Kaul19855}
\bibinfo{author}{\bibfnamefont{S.}~\bibnamefont{Kaul}}, \bibinfo{journal}{J.
  Magn. Magn. Mater.} \textbf{\bibinfo{volume}{53}}, \bibinfo{pages}{5 }
  (\bibinfo{year}{1985}).

\bibitem[{\citenamefont{Jiang and Chien}(1996)}]{jiang:5615}
\bibinfo{author}{\bibfnamefont{J.~S.} \bibnamefont{Jiang}} \bibnamefont{and}
  \bibinfo{author}{\bibfnamefont{C.~L.} \bibnamefont{Chien}},
  \bibinfo{journal}{J. Appl. Phys.} \textbf{\bibinfo{volume}{79}},
  \bibinfo{pages}{5615} (\bibinfo{year}{1996}).

\bibitem[{\citenamefont{Mohan et~al.}(1998)\citenamefont{Mohan, Kronm\"uller,
  and Kelsch}}]{PhysRevB.57.2701}
\bibinfo{author}{\bibfnamefont{C.~V.} \bibnamefont{Mohan}},
  \bibinfo{author}{\bibfnamefont{H.}~\bibnamefont{Kronm\"uller}},
  \bibnamefont{and} \bibinfo{author}{\bibfnamefont{M.}~\bibnamefont{Kelsch}},
  \bibinfo{journal}{Phys. Rev. B} \textbf{\bibinfo{volume}{57}},
  \bibinfo{pages}{2701} (\bibinfo{year}{1998}).

\bibitem[{\citenamefont{Choi et~al.}(2004)\citenamefont{Choi, Haskel, Camley,
  Lee, Lang, Srajer, Jiang, and Bader}}]{PhysRevB.70.134420}
\bibinfo{author}{\bibfnamefont{Y.}~\bibnamefont{Choi}},
  \bibinfo{author}{\bibfnamefont{D.}~\bibnamefont{Haskel}},
  \bibinfo{author}{\bibfnamefont{R.~E.} \bibnamefont{Camley}},
  \bibinfo{author}{\bibfnamefont{D.~R.} \bibnamefont{Lee}},
  \bibinfo{author}{\bibfnamefont{J.~C.} \bibnamefont{Lang}},
  \bibinfo{author}{\bibfnamefont{G.}~\bibnamefont{Srajer}},
  \bibinfo{author}{\bibfnamefont{J.~S.} \bibnamefont{Jiang}}, \bibnamefont{and}
  \bibinfo{author}{\bibfnamefont{S.~D.} \bibnamefont{Bader}},
  \bibinfo{journal}{Phys. Rev. B} \textbf{\bibinfo{volume}{70}},
  \bibinfo{pages}{134420} (\bibinfo{year}{2004}).

\bibitem[{\citenamefont{Kirby et~al.}(2009)\citenamefont{Kirby, Watson, Davies,
  Zimanyi, Liu, Shull, and Borchers}}]{kirby:07C929}
\bibinfo{author}{\bibfnamefont{B.~J.} \bibnamefont{Kirby}},
  \bibinfo{author}{\bibfnamefont{S.~M.} \bibnamefont{Watson}},
  \bibinfo{author}{\bibfnamefont{J.~E.} \bibnamefont{Davies}},
  \bibinfo{author}{\bibfnamefont{G.~T.} \bibnamefont{Zimanyi}},
  \bibinfo{author}{\bibfnamefont{K.}~\bibnamefont{Liu}},
  \bibinfo{author}{\bibfnamefont{R.~D.} \bibnamefont{Shull}}, \bibnamefont{and}
  \bibinfo{author}{\bibfnamefont{J.~A.} \bibnamefont{Borchers}},
  \bibinfo{journal}{J. Appl. Phys.} \textbf{\bibinfo{volume}{105}},
  \bibinfo{eid}{07C929} (\bibinfo{year}{2009}).

\bibitem[{\citenamefont{Kirby et~al.}(2006)\citenamefont{Kirby, Borchers,
  Rhyne, O'Donovan, te~Velthuis, Roy, Sanchez-Hanke, Wojtowicz, Liu, Lim
  et~al.}}]{kirby:245304}
\bibinfo{author}{\bibfnamefont{B.~J.} \bibnamefont{Kirby}},
  \bibinfo{author}{\bibfnamefont{J.~A.} \bibnamefont{Borchers}},
  \bibinfo{author}{\bibfnamefont{J.~J.} \bibnamefont{Rhyne}},
  \bibinfo{author}{\bibfnamefont{K.~V.} \bibnamefont{O'Donovan}},
  \bibinfo{author}{\bibfnamefont{S.~G.~E.} \bibnamefont{te~Velthuis}},
  \bibinfo{author}{\bibfnamefont{S.}~\bibnamefont{Roy}},
  \bibinfo{author}{\bibfnamefont{C.}~\bibnamefont{Sanchez-Hanke}},
  \bibinfo{author}{\bibfnamefont{T.}~\bibnamefont{Wojtowicz}},
  \bibinfo{author}{\bibfnamefont{X.}~\bibnamefont{Liu}},
  \bibinfo{author}{\bibfnamefont{W.~L.} \bibnamefont{Lim}},
  \bibnamefont{et~al.}, \bibinfo{journal}{Phys. Rev. B}
  \textbf{\bibinfo{volume}{74}}, \bibinfo{eid}{245304} (\bibinfo{year}{2006}).

\end{thebibliography}
\providecommand{\noopsort}[1]{}\providecommand{\singleletter}[1]{#1}%

\end{document}